\documentstyle[11pt]{article}
\begin{document}
\vbox{
\begin{flushright}
AZPH-TH/93-32
\end{flushright}
\title{THE  PARTONIC CONTENT OF $h_1(x)$ AND $h_2(x)$}
\author{Arjun Berera** \\
Department of Physics \\
University of Arizona \\
Tucson, Arizona 85721}
\date{}
\maketitle
\begin{abstract}
\baselineskip .45 in
A light-cone wavefunction interpretation is presented for the polarized
distribution functions $h_1(x)$ and $h_2(x)$.  All matrix elements
for moments of these distributions are given in terms of overlap
integrals between Fock state amplitudes of the target state.
In a suitable spinor basis, $h_1(x)$ involves only diagonal
matrix elements so can be interpreted as a density.  Matrix
elements of $h_2(x)$ connect Fock states differing
by one gluon so that $h_2(x)$ has no simple interpretation as a density.
Nevertheless, in the wavefunction decomposition, $h_2(x)$ is described
through a compact set of elementary quark-gluon processes
which are averaged over the target wavefunction.
\end{abstract}
\vspace{20mm}}

\bigskip
\eject
%\noindent {\Large\bf I. Introduction}
\medskip

In a theoretical study of the polarized Drell-Yan process,
Ralston and Soper \cite{ralston} showed that the cross section
for polarized $p-\bar p$ scattering can be expressed
in terms of nine structure functions.  From their analysis,
they derived a formula in polarized Drell-Yan production
between transversely polarized $p\bar p$ pairs for the spin
asymmetry defined as,
$$
A_{\lambda _{A} \lambda _{B}} = {\sigma (\lambda _A,\lambda _B)
- \sigma (\lambda _A,-\lambda _B)
\over \sigma (\lambda _A,\lambda _B) + \sigma (\lambda _A,-\lambda _B)}
\eqno{(1-1)}
$$
where $\lambda_i$ is the polarization of particle i
and $-\lambda_i$ means the opposite polarization for the same particle.
Their relation was expressed in terms of the distribution
$h_1(x)$, often referred to as the transversity distribution,
to be,
$$
A_{T T} = {\sin ^2 \theta \cos 2 \phi \over 1 + \cos ^2 \theta} \
{\sum_a e_a^2 \  h_1^a(x) \ h_1^a(y) \over \sum_a e_a^2 \ f_1^a (x) \ f_1^a(y)}
\ , \eqno{(1-2)}
$$
where $\phi$ is the azimuthal angle and x and y are the longitudinal
momentum fractions carried by the quarks in the two impinging
hadrons.
In following these lines, Jaffe and Ji \cite{jaffe} obtained
the spin asymmetry for longitudinal-transverse collisions
in terms of $h_2(x)$ to be,
$$
A_{LT} = {2\sin 2\theta \cos \phi \over 1 + \cos ^2\theta} \
{M \over Q} \  {\sum_a e_a^2\ [g_1^a(x) y g_{T}^l(y)+xh_L^a(x)\ h_1^a(y)] \over
\sum_a e^2_a\ f_1^a(x)\ f_1^a(y)}
\eqno{(1-3)}
$$
where,
$$
h_L(x) \equiv h_1(x) + {h_2(x) \over 2}.
$$
Since their introduction, for the most part these distributions
have remained relatively
unmentioned within the standard lines.  However with the advent of
a polarized proton-proton beam at Argonne and future hopes of
higher energy polarized proton beams, these distributions
are reaching a stage of tangibility.  On the theoretical side,
there has been little theoretical development of them.  A significant amount
of clarification was given to them by Jaffe and Ji \cite{jaffe}
who illuminated their operator product tensor structure.
To reach a closer physical connection, what remains is
still to analyze the parton wavefunction interpretation for them.
It is the purpose of this paper to do this.

In this paper we will study the light-cone wavefunction interpretation
of $h_1(x)$ and $h_2(x)$.  We will derive explicit formulas
for the moments of these distributions
in terms of
parton correlations.
For $h_1(x)$ such an undertaking is less significant since it
has an interpretation as a quark-antiquark density in a suitable basis.
On the other hand, $h_2(x)$ has no such simple interpretation.
For this reason its physical meaning has remained somewhat obscure
However, as we shall see, its parton content can be understood
through only six types of matrix elements as opposed
to two for $h_1(x)$.

The paper is organized as follows.  In section two we review
the operator-product-expansions for the moments of $h_1(x)$ and $h_2(x)$.
In section three we discuss the light-cone wavefunction formalism
in a manner suitable for our further developments.  In section four
we give our light-cone wavefunction analysis of $h_1(x)$ and $h_2(x)$.
Finally we give some closing comments and discuss future directions in
the conclusion.  There is also an appendix which tabulates various
light-cone spinors as a reference since we often found them to be useful
in the analysis.

\bigskip
\noindent {\Large\bf Section 2}
\medskip

In this section we will review the formal properties of
$h_1$and $h_2$ based on the work of Ralston and Soper \cite{ralston}
and Jaffe and Ji\cite{jaffe}.  These distributions are defined
through matrix elements between the
target proton state of a specific quark bilinear operator
as,
\begin{eqnarray*}
{\int{d\lambda\over 2\pi}e^{i\lambda x}\langle PS|\bar\psi(0)
\sigma_{\mu\nu}i\gamma_5\psi(\lambda n)|PS\rangle} & \equiv
&{2 \left[ h_1(x)(S_{\perp\mu}p_\nu-S_{\nu\perp} p_\mu)/M\right.} \\
 & & {+(h_1(x)+{{h_2(x)} \over 2})M(p_\mu n_\nu-p_\nu n_\mu)(S\cdot n)} \\
 & & {\left. +h_3(x)M(S_{\perp \mu}n_\nu-S_{\perp \nu}n_\mu)\right] \ . }
\end{eqnarray*}
$$\eqno (2-1)$$
where $n^2=n^+=0$.
By a suitable choice of external spin and momenta, the desired densities
can be isolated as,
$$
h_1(x)={1\over\sqrt2 p^+}\int{d\lambda\over2\pi}e^{i\lambda x}\langle PS_\perp
|\psi^{\hbox{\dag}}_+(0)\gamma_\perp\gamma_5\psi_+(\lambda n)|PS_\perp
\rangle \ . \eqno(2-2)
$$
which is the twist two component and,
$$
(h_1(x)+{{h_2(x)} \over 2})\equiv
h_L(x)={1\over2M}\int{d\lambda\over2\pi}e^{i\lambda x}\langle PS_z
|\psi^{\hbox{\dag}}_-(0)\gamma_0\gamma_5\psi_+(\lambda n)-
\psi^{\hbox{\dag}}_+(0)\gamma_0\gamma_5\psi_-(\lambda n)|PS_z\rangle
\ . \eqno(2-2b)
$$
which contains a mixture
of both twist two and three components.
Recall that in the light-cone formalism
of \cite{kogut},
the twist is directly related to the number of good and bad fields
in the quark bilinear with twist two have no bad components
and twist three having one.

To obtain the operator product expansion for the matrix elements
above we must perform a covariant Taylor expansion of the
quark field about the origin on the left hand side in (2-1) and compare
like terms to the right hand side in order to extract
the appropriate bilinear operator expansion terms.
For $h_1(x)$ the relevant operator bilinear
one obtains is \cite{jaffe},
$$
\theta^{\nu\mu_1 \ \mu_2\cdots\mu_n}\equiv S_n\bar\psi
i\gamma_5\sigma^{\nu\mu_1}
iD^{\mu_2}\cdots iD^{\mu_n}\psi \ . \eqno(2-3)
$$
Computing its matrix element between the proton state
one gets,
$$
 \langle PS|\theta^{\nu\mu_1 \ \mu_2\cdots\mu_n}|PS\rangle
=2a_nS_n(S^\nu P^{\mu_1}
-S^{\mu_1}P^\nu)P^{\mu_2}\cdots P^{\mu_n}/M + {\rm \ terms \ involving \ }
g^{\mu_i\mu_j} \ . \eqno(2-4)
$$
where $a_n$ is related to $h_1(x)$ by the moments relation,
$$
\int^\infty_{-\infty}dx \ x^{n-1}h_1(x)=\int^1_0 dx \ x^{n-1}(h_1(x)-
(-1)^{n-1} \bar h_1(x))=a_n \ . \eqno(2-5)
$$
and $\bar h_1(x)$ is defined like (2-2) except for antiquarks.

For $h_2(x)$ we consider the Jaffe-Ji equivalent to the Wilczek-Wadzuda
decomposition of $g_2(x)$ and write,
$$
h_L(x)=2x\int^1_x{h_1(y)\over y^2}dy+{m\over M}\left[{g_1(x)\over x}-x\int^1_0
{g_1(y)\over y^2}dy\right]+h^3_L(x) \ . \eqno(2-6)
$$
where the moments of $h^3_L(x)$ are,
$$
{\cal M}_n[h^3_L]=-\sum^{[(n+1)/2]}_{l=2}\left(1-{2l\over n+2}\right)b_{n,l}
\ . \eqno(2-7)
$$
with,
$$
b_{n,l}\equiv c_{n,n-l}-c_{n,l}
\ . \eqno(2-8)
$$
here $c_{n,l}$ is obtained from the matrix element,
$$
\langle
PS|\theta^{\mu_1\cdots\mu_n}_l|PS\rangle\equiv2c_{nl}MS_nS^{\mu_1}P^{\mu_2}
\cdots P^{\mu_n}  \eqno(2-9)
$$
with
$$
\theta_{l}^{\mu_1\cdots\mu_n}={1\over2}S_n\bar\psi\sigma^{\alpha
\mu_1}i\gamma_5iD^{\mu_2}
\cdots igF^{\mu_l}_\alpha
\cdots iD^{\mu_n}\psi \ \hbox{-- \ traces} \ . \eqno(2-10)
$$
\bigskip
\bigskip
\noindent {\Large\bf Section 3}
\medskip

In this section we will review salient features of light-cone
field theory that will be relevant for our calculation.
The reader is urged to also examine reference \cite{mankiewicz}
which supplements the discussion in this section.  Our purpose here is
to explicate necessary quantities used in our work
for the convenience of
the reader.

We imagine a particle which in its rest frame has its spin quantization
axis defined along the z-direction.  The particle is now observed
from a frame moving at the idealized
limit of the speed of light in the -z-direction.
We now desire to describe the kinematic
properties of this particle , be it quarks or gluons, in
terms of light-cone coordinates of this boosted
frame and in the light-cone gauge.  The
light-cone coordinates (no subscript) are defined with respect to the
spacetime coordinates, {$x_s$}, as,
\begin{eqnarray*}
x^+ &=&x_-={x^0_s+x^3_s\over\sqrt2} \\
x^- &=&x_+={x^0_s+x^3_s\over\sqrt2} \\
x^\perp &=&x^\perp_s \ .
\end{eqnarray*}
$$
\eqno (3-1)
$$
where $x^+$ is taken as the light-cone time and
the scalar product of two four-vectors $v_1$and $v_2$ is
$v_1\cdot v_2=v_1^+v_2^-+v_1^-v_2^+-v_1^\perp\cdot v_2^\perp$.

We introduce the fields and definitions needed in our work through
the QCD Lagrangian,
$$
{\cal L}=-{1\over2} Tr(F^{\mu\nu}F_{\mu\nu})+\bar
\psi(i{\cal D}^{\mu}-m)\psi \eqno(3-2)
$$
where the covariant derivative $i{\cal D}^{\mu}=i{\cal \partial}^{\mu}-g{\cal
A}^{\mu}$
and field strength
$F^{\mu\nu}= \partial^{\mu}A^{\upsilon}-\partial^{\upsilon}A^{\mu}$.
The quark field $\psi$ is a color triplet spinor where only one
flavor is written above.  The gluon gauge field $A^{\mu}$=
$\sum_aA^{a\mu}T^{a}$
 is a traceless 3x3 color matrix.  The independent
dynamical fields at a given light-cone time,
say $x^+$=0, are $\psi_+=\Lambda_+\psi$ and $A^i$ (i=1,2),
where the projection operators $\Lambda_{\pm} \equiv {1 \over
2}\gamma^0\gamma^{\pm}$
with $\gamma^{\pm} \equiv \gamma^0 \pm \gamma^3$
and $\partial^{\pm}=\partial^0=-\partial^3$.  From the
equations of motion, the remaining fields can be obtained.
With $\psi=\psi_++\psi_-$, one can express $\psi_-$
in terms of $\psi_+$ as,
$$
\psi_-\equiv\Lambda_-\psi={1\over i\partial^+}[i\vec D_\perp\cdot\vec
\alpha_\perp
+\beta m]\psi_+=\tilde\psi_--{1\over i\partial^+}g\vec A_\perp\cdot\vec a_\perp
\psi_\perp
\ . \eqno(3-3)
$$
Turning to the gauge field in the light-cone gauge
one has,
$$
A^+=0 \eqno(3-4a)
$$
$$
A^-={2\over i\partial^+}i\vec\partial_\perp\cdot\vec A_\perp+
{2g\over (i\partial^+)^2}\{[i\partial^+A^i_\perp,A^i_\perp]+2
\psi^{\hbox{\dag}}_+T^a\psi_+T^a\} \eqno(3-4b)
$$
with $\beta=\gamma^0$ and $\vec \alpha_\perp=\gamma^0 \vec
\gamma_\perp$.
Upon quantization, the independent dynamical coordinates
can be decomposed into creation and annihilation operators
at a given light-cone time, say $x^+=0$ again, as,
\begin{eqnarray*}
\psi_+(x)&=&\int_{k^+>0}{dk^+d^2k\over k^+16\pi^3}\sum_\lambda
\{b({k^+,\vec k},
\lambda) u_+(k^+,{\vec k},\lambda)e^{-ik\cdot x}+d^{\hbox{\dag}}
(k^+,{\vec k},\lambda)v_+(k^+,{\vec k},\lambda)e^{ik\cdot x}\}_{\tau=x^+=0} \\
A^i_\perp(x)&=&\int_{k^+>0}{dk^+d^2k\over k^+16\pi^3}\sum_\lambda\{a
(k^+,{\vec k},\lambda)\epsilon^i_\perp(\lambda)e^{-ik\cdot x}+c.c.
\}_{\tau=x^+=0} \ ,
\end{eqnarray*}
$$ \eqno{(3-5)} $$
with commutation relations,
\begin{eqnarray*}
\{b({k^+,\vec k},\lambda),\ b^{\hbox{\dag}}({p^+,\vec p},\lambda)\}
&=&\{d({k^+,\vec k_\perp},\lambda), \ d^{\hbox{\dag}}({p^+,\vec p_\perp},
	\lambda^\prime)\} \\
&=&[a({k^+,\vec k_\perp},\lambda), \
a^{\hbox{\dag}}({p^+,\vec p_\perp},\lambda)] \\
&=&16\pi^3k^+\delta(k^{+}-p^{+})\delta^{(2)}(\vec k_\perp - \vec p_\perp)
 \delta_{\lambda \lambda^\prime}
\end{eqnarray*}
$$
\eqno (3-6a)
$$
$$\{b,b\}=\{d,d\} = \cdots=0 \ . \eqno(3-6b)
$$
In calculation, it is often more convenient to treat the fermion field not
separately in terms of its plus and minus components but rather to write
$\psi_-(x)$ as,
$$
\psi_-(x)=\Phi_-(x)-{g\over i\partial^+}(\alpha_\perp A_\perp\Psi_\perp)
\ , \eqno(3-7)
$$
where
$$
\Phi_-(x)=\left({1\over i\partial^+}\right)[(-i\alpha_\perp\partial_\perp+
\beta m)\psi_+] \ . \eqno(3-8)
$$
Next defining
$\psi(x)=\psi_{-}(x)+\psi_{+}(x)$,
its expansion at $x^{+}=0$ is,
$$
\psi(x)= \int_{k^{+}>0}{{dk^+d^{2}k_\perp}\over{16\pi^{3}k^{+}}}
\sum_{\lambda=_{-}^{+}1/2}[b(k^+, k_\perp, \lambda)u(k^+,k_\perp, \lambda)
e^{-ikx} + d^{+}(k^+, k_\perp, \lambda)v(k^+, k_\perp,
\lambda)e^{ikx}]_{x^{+}=0}
$$
$$
\eqno{(3-9)}
$$
Here $u,v$ are light-cone spinors which are explicitly  given in the
Appendix, Eqs (A-1).

Having defined the field theory, let us now turn to the Fock space
representation of the wavefunctions which we will sometimes
refer to as hadronic wavefunctions.
For our purpose, we can consider the wavefunction as a expansion
in the Fock space at some renormalization scale $\mu$.
We will not be addressing issues regarding zero modes
or ground state properties nor matters of renormalization.
Our concern is only to examine the Fock space correlations contained in
the distributions $h_1(x)$ and $h_2(x)$.

The light-cone hadron wavefunction can be thought of
as an infinite dimensional column vector representation for Fock
space.  The Fock space of interest here is of the many body quark
and gluon system.  Each Fock space component is
a product of quark and gluon quanta with specified
quantum numbers.  Premultiplying this component is a complex valued
factor which specifies the probability amplitude to find that
state in the normalized hadronic wavefunction.  A qualitative
remark about the light-cone wavefunction formalism is that
scattering processes have static versus dynamic description
in that it is more natural here to imagine constructing
the bound state wavefunction first, with in particular
its high energy tail, and then coupling
it to a scattering process.  This contrasts the spacetime
picture in which the "building" of the wavefunction
is often incorporated within the Feymann diagram representing
the scattering process.  A useful example is the Alterilli
Parisi evolution equations
for deep-inelastic-scattering
which are typically interpreted as a radiative
bremstalung by the parton before scattering.  Alternatively,
if one thinks of the parton as part of a bound state system,
which it is, this same process of radiation is
actually generating the high energy Fock space components of
the wavefunction.  Mathematically, of course , both methods
are equivalent, however each has its advantages in analyzing
physical processes.  The light-cone wavefunction approach
provides a useful language for the analysis of partonic
correlations in hadrons and so extends naturally
for use in modeling low energy hadronic structure.

Turning to the representation of the light-cone
wavefunction we write it as,
%------------------------------
%changed to an equation array to prevent overflow.
%$$
%|PS\rangle=\sum^\infty_{n=qqq}\sum_{|f_1,\ldots,f_n|}\sum_{|\lambda_1,\ldots,
%\lambda_n|}\int{[dx]_n[d^2k_1]_n\over \sqrt{x_1\cdots x_n}}\psi^n_S
%(k_{\perp i},x_i,\lambda_i)a\hbox{\dag}(f_1,p_1^+, p_{\perp
% 1},\lambda_1)\cdots
%a\hbox{\dag}(f_n,p_n^+, p_{\perp n},\lambda_n|\Omega\rangle \ , \eqno(3-10)
%$$
%---------------------------------
\begin{eqnarray*}
{|PS\rangle}&{=}&{\sum^\infty_{n=qqq}\sum_{|f_1,\ldots,f_n|}\sum_{|\lambda_1,\ldots, \lambda_n|}\int{[dx]_n[d^2k_1]_n\over \sqrt{x_1\cdots x_n}} \psi^n_S
(k_{\perp i},x_i,\lambda_i)a\hbox{\dag}(f_1,p_1^+, p_{\perp 1},\lambda_1)} \\
&{\cdots}&{a\hbox{\dag}(f_n,p_n^+, p_{\perp n},\lambda_n | \Omega)}
\end{eqnarray*}
$$
\eqno (3-10)
$$
where
$$
[dx]_n=\prod^n_{i=1}dx_i\delta\left[\sum^n_{i=1}x_i-1\right] \ ,
$$
$$[d^2k_\perp]_n=\prod^n_{i=1}{d^2k_{{\perp}i}\over16\pi^3}\delta^{(2)}\left[\sum^n_{i=1}k_{\perp i}
\right] \ .
$$

\bigskip
\noindent {\Large\bf Section 4}
\medskip

Up to now, everything we have said is well known.  Now we will apply
this methodology to compute light-cone wavefunction matrix elements
of $h_1(x)$ and $h_2(x)$.
For any non-singlet SU(3)-flavor channel, one must insert
the appropriate Gell-Mann $\lambda$-matrix between the fermion fields.
In the singlet-channel, the quark and gluon fields mix under
renormalization.  This requires further consideration in order
to make a partonic interpretation.  We will not
pursue this matter in this work.
This section is divided into two parts,
In the first part, we will examine $h_1(x)$, which entails no
complications since in an appropriate basis its matrix elements
are all diagonal.  In the second part we will examine $h_2(x)$, for
which a detailed analysis is required of the relevant fermion bilinear operator
and the diagonal and nondiagonal matrix elements it yields.
Nevertheless, once the dust settles, we find that the final answer is
reasonably simple.

For a general operator, $\theta^{[\mu]}$,
we are interested in its matrix element
between the hadron wavefunction.  In general this matrix element
decomposes into a sum of terms, each one connecting a particular
bra Fock space state to one of the ket states.
Our task is to evaluate all the individual contributions
arising from the Fock space decomposition
for the matrix elements of (2-2a) and (2-2b).
It is convenient to factor each of the individual terms
into two pieces, which we refer to as
the momentum and spin factors. The momentum factor,
$f_p(q,q';\{g\}_k,\{g\}_{k'})$,
contains the momentum dependent contributions
of the covariant derivatives when acting on
the two fermions, q and q', and n-2 gluons, {g}, for twist n
that are being connected by the operator.
The spin factor $O_s (q,q';\{g\}_k,\{g\}_{k'})$, on the other hand,
contains the contribution from the fermion bilinear operator
and gluon polarization operators when acting
on the connected states.
It depends on the two momentum and spin states of the connected
fermions and the momentum and polarization states of the connected
gluons.

With this decomposition, we can write the general form of the matrix
element as,
$$
\langle PS | \theta^{\mu} | PS \rangle = \sum_
{\begin{array}{c}
{q \epsilon \{ \bar{f},f \}_{n}} \\
{q^{'} \epsilon \{ \bar{f}, f \}_{{n}'}}
\end{array}}
\sum_
{\begin{array}{l}
{\{g\}_k \epsilon \{g\}_m } \\
{\{g'\}_{k'} \epsilon \{g'\}_{m'}}
\end{array}}
\sum_{\{\lambda\}} \  \eqno(4-1)
$$
\bigskip
\begin{eqnarray*}
& {\int [[dx_q d^{2}k_{\perp q} dx_{q'} d^{2}k_{\perp q'} (d^{3}k_g)_k
(d^{3}k_{g'})_{k'}} \\
& {(d^{3}k_q)_{n-1} (d^{3}k_{q'})_{n'-1} (d^{3}k_{g})_{m-k}
(d^{3}k_{g'})_{m'-k'}]]} \\
& {f_p(q, q', \{g\}_k, \{g\}_{k'})} \\
& {\Psi^{mn*}_{s}(x_{q_1}, k_{\perp {q_1}}, \lambda_{q_1},
\ldots, x_q, k_{\perp q},} \\
& {\lambda_q, \ldots, x_{q_n}, k_{\perp q_n}, \lambda_{q_n}; \{g\}_m)} \\
& {O_s(q, q', \{g\}_k, \{g\}_{k'})} \\
& {\Psi^{m'n'}_{s}(x_{q'_1}, k_{\perp {q'_1}}, \lambda_{q'_1}, \ldots, x_{q'},}
\\
& {k_{\perp q'}, \lambda_{q'}, \ldots, x_{q'_{n'}}, k_{\perp q'_{n'}},
 \lambda_{q'_{n'}},; \{g'\}_{m'})}
\end{eqnarray*}
where $\theta^{[\mu]}$ in general is a quark bilinear operator
with tensor indices [$\mu$].  Above, the double square brackets around
all the momentum differentials are to imply insertion of a single
$\delta$-function factor for each the bra and ket states of the form,
$$
\delta\left[\sum^{n+m}_{i=1}x_i-1\right]
\delta^{(2)}\left[\sum^{n+m}_{i=1}k_{\perp i}
\right]
\delta\left[\sum^{n'+m'}_{i=1}x'_i-1\right]
\delta^{(2)}\left[\sum^{n'+m'}_{i=1}k
^\prime_{\perp i}
\right]
$$
Also for notational brevity,
$$
(d^3k)_n \equiv
\prod^n_{i=1}{{dx_i}\over{\sqrt{x_i}}}{{d^2k_{{\perp}i}}\over{16 \pi^3}},
$$
\noindent {\Large\bf $h_1(x)$}

To compute the coefficient $a_n$ in (2-4) for $h_1(x)$, it is sufficient
to evaluate the operator $\theta^{[\mu]}$,
with all the symmetrized indices $\mu_1\mu_2$-$\mu_n$
carrying Lorentz index + and the remain index $\nu$ carrying a transverse
component which we take as x.  For the proton state, it is convenient to
observe it in its center of mass frame where
$P^+_{cm}$=$P^-_{cm}={M\over{\sqrt{2}}}$ .  Boost
invariance in the light-cone allows us to make such a choice.   For
the polarization state of the proton, consistent with our above choice
we take it polarized in the +x-direction so that S=(0,0,M,0).
The coefficient $a_n$ can then be obtained from
$$
M_{+x}\equiv\langle PS_{+x}|
-\bar\psi {\gamma^x\gamma^+\gamma^5}
(i\partial^+)^{n-1}\psi|PS_{+x}\rangle = 2a_nS^x(P^+)^n/M
\eqno (4-2)
$$
Defining as in \cite{jaffe},
$$
Q^i_\pm\equiv{1\over2}(1\mp\gamma^5\gamma^i)
\eqno(4-3)
$$
with i=1,2 being the transverse coordinates x and y respectively,
we find,
$$
Q^{i}_{\pm}u^{i}_{+}(k, \lambda=\pm)=u^{i}_{+}(k, \lambda=\pm)
$$
$$
Q^{i}_{\pm}v^{i}_{+}(k, \lambda=\pm)=v^{i}_{+}(k, \lambda=\pm)
$$
$$
Q^{i}_{\pm}u^{i}_{+}(k, \lambda=\stackrel{-}{+})=Q^{i}_{\pm}v^{i}_{+}
(k, \lambda=\stackrel{-}{+})=0
\eqno(4-4)
$$
where the "transverse spinors" $u^i_{\pm}, v^i_{\pm} $ are given in the
appendix eqs (A-2).
Reexpressing (4-2) in terms of (4-3) we obtain,
%--------------------------------------------------------------------------
%working eqn h1-2-5 begins here
\begin{eqnarray*}
%---------------------------------------------------------------------------
{2a_n(P^+)^n}&{=}&{\sum^\infty_{m=qqq}
\sum^{\pm{1/2}}_{[\lambda_1\cdots\lambda_m]}
\sum^{\infty}_{p=0}\sum_{[\lambda_g]_p}\left\{ \sum_{s,s' \epsilon \{f\}_m}
\int [d^3k_{(f,\bar f)}d^3k_g]_{(m+p)}(P^+)^{n-1} \right. } \\
%---------------------------------------------------------------------------
&{\times}&{\psi^{mp*}_{S_{+x}}(x_1,k_\perp,\lambda_1,\ldots x_{s^\prime},
k_{\perp {s^\prime}},\lambda_{s^\prime}\cdots x_m,k_{\perp
m},\lambda_m;\{g\}_p)}\\
%---------------------------------------------------------------------------
&{\times}&{\left[x^n_s{\bar u^1_+(\lambda_{s^\prime})\over\sqrt{x_{s^\prime}}}
(Q^1_+
-Q^1_-)\gamma^+{u^1_+(\lambda_s)\over\sqrt{x_s}}\right]\psi^{mp}_{S_{+x}}
(x_1,k_{\perp1},\lambda_1,\ldots,x_{s},k_{\perp
s},\lambda_s,\ldots,x_m,k_{\perp m},\lambda_s;\{g\}_p)} \\
%---------------------------------------------------------------------------
& {+} & {\sum_{s,s' \epsilon \{ \bar f\}_m}
\int[d^3k_{(f,\bar f)}d^3k_g]_{(m+p)}(P^+)^{n-1}(-1)^{n}} \\
%---------------------------------------------------------------------------
&{\times}&{\psi^{mp\ast}_{S_{+x}}
(x_1, k_{\perp 1},\lambda_1
\cdots x_{s^\prime},k_{\perp {s^\prime}},\lambda_{s^\prime},
\cdots x_m,k_{\perp m},\lambda_m;\{g\}_p)} \\
%---------------------------------------------------------------------------
%& {\times} & \left. {\left[x^n_s \frac{\bar{v}^1_{\lambda^{\prime}_{s}}}
%{\sqrt{x{s^\prime}}(Q^1_+ -Q^1_-) \frac{v^1_{\lambda_s}}{\sqrt{x_s}}}\right]
%\psi^{mp}_{S_{+x}}
%(x_1,k_{\perp1},\lambda_1\ldots x_s,k_{\perp s},\lambda_s,\ldots,
%x_m,k_{\perp m} \lambda_m)} \right\}
& {\times} & \left.
{\left[ x^n_s
{\bar{v}^1_+(\lambda_{s}^{\prime}) \over \sqrt{x_{s^{\prime}}}}
(Q^1_+ -Q^1_-)
{v^1_+(\lambda _s)\over \sqrt{x_s}} \right]
\psi^{mp}_{S_{+x}}
(x_1,k_{\perp1},\lambda_1\ldots x_s,k_{\perp s},\lambda_s,\ldots,
x_m,k_{\perp m} \lambda_m)} \right\}
%---------------------------------------------------------------------------
\end{eqnarray*}
$$
\eqno(4-5)
$$
where the sum on m is to imply both quarks and antiquarks and
one factor of -1 in the second term is due to the ordering
of antiquarks.  Recall also that a Fock space amplitude coefficient
for n-quanta,
$\psi^n(\{k\})$, has mass dimensions $-$(n-1).  One can check by inspection
above that $a_n$ is dimensionless.  All the contributions to the matrix
element for $a_n$ in this basis, where
the quark spins
are along the x-direction,
are now
between the same Fock space state with
no mixing between states.
In terms of the general form (3-10) we can identify
$f_p(q,q';\{g\}_k,\{g\}_{k'})$ as,
$$
f_p(q,q';\{g\}_k,\{g\}_{k'})=
(P^+x_q)^n=({M\over{\sqrt{2}}}x_q)^n \eqno(4-6)
$$
To calculate the spin function
$O_s (q,q';\{g\}_k,\{g\}_{k'})$,
we evaluate,
$$
{\bar u^1_+({\lambda {s^\prime}})\over\sqrt{x_{s'}}}(Q^1_+-Q^1_-)\gamma^+
{u^1_+({\lambda s})\over\sqrt{x_s}}
\eqno(4-7a)
$$
for quarks, and
$$
(-1)^{n}{\bar v^1_+({\lambda_{s^\prime}})\over\sqrt{x_{s'}}}(Q^1_+-Q^1_-)
\gamma^+{v^1_+({\lambda_{s}})\over\sqrt{x_s}}
\eqno(4-7b)
$$
for antiquarks, which leads to the expectation values,
$$
P^+\langle\delta_{\lambda_{s^\prime_x\uparrow}
\lambda_{s_x\uparrow}}-\delta_{{\lambda_{s^\prime_x\downarrow}}
\lambda_{s_x\downarrow}}\rangle
\eqno(4-8a)
$$
and
$$
P^+\langle\delta_{\lambda_{s^\prime_x\downarrow}
\lambda_{s_x\downarrow}}-\delta_{{\lambda_{s^\prime_x\uparrow}}
\lambda_{s_x\uparrow}}\rangle
\eqno(4-8b)
$$
respectively.
%-----------------
We have now given the explicit form of all matrix elements
of $h_1(x)$.  In regards to our earlier discussion, in this basis
we can readily identify $h_1(x)$ as a density since the
nonvanishing contributions to the matrix elements needed to compute
its moments do not mix Fock space components.
Explicitly we can express the moments of $h_1(x)$
as,
$$
\int^1_0 dx x^{n-1} (h_1(x)-(-1)^{n-1}\hbar_1(x)) = a_n =
\int^1_0 x^{(n-2)}[T_q(x)-(-1)^{(n-1)}T_{\bar q}(x)]  \  \ n=1,2,\cdots
\eqno(4-9)
$$
where
%----------------------------------------------------------------------------
%this was so wrong i commented it out and redid it.
%\begin{eqnarray*}
%\delta^{(2)}(P-P^prime)
%{T_{q(\bar{q})}(x)} & { = } & {\int{d^2k_\perp\over16\pi^2}\langle PS_\perp |
%b^{+}_{q(\bar{q})}(xP,k_{\perp},\uparrow)
% b_{q(\bar{q})}(xP,k_\perp,\uparrow)-b^{+}_{q(\bar{q})}
%(xP,k_{\perp},\downarrow)b_{q(\bar{q})}(xP,k_{\perp},\downarrow) |
% P^\primeS_{\perp}^\prime
%\rangle \\
% & { = } & {\sum^\infty_{m=0}\sum^\infty_{n=3}
%\int[d^3k_g]_m\int_{n}\hspace{-13pt} \Sigma
%[|\Psi^{nm}_{S_{+x}}
%(x_{1}, k_{\perp 1}, \lambda_1, \ldots, x_{n_\bar{q}}, k_{\perp n_\bar{q}},
%\lambda_{n_\bar{q}},x_{n_\bar{q}+1},k_{\perp n_\bar{q} +1},
%\lambda_{n_\bar{q}+1},
%\ldots, x_n, k_{\perp n}, \lambda_{n}j\{g\})|^{2}}
%\end{eqnarray*}
%----------------------------------------------------------------------
\begin{eqnarray*}
{\delta^{(2)}(P_\perp-P_\perp{^\prime}) T_{q(\bar{q})}(x)} & { = } &
{\int{d^2k_\perp\over16\pi^3}\langle PS_\perp |
b^{+}_{q(\bar{q})}(xP,k_{\perp},\uparrow)
b_{q(\bar{q})}(xP,k_\perp,\uparrow)} \\
&{-}&{b^{+}_{q(\bar{q})}
(xP,k_{\perp},\downarrow)b_{q(\bar{q})}(xP,k_{\perp},\downarrow) |
P^\prime S_{\perp}
\rangle}
\end{eqnarray*}
$$ \eqno{(4-10a)} $$
\noindent which gives,
\begin{eqnarray*}
T_{q(\bar{q})}(x) & { = } & {\sum^\infty_{m=0}\sum^\infty_{n=3}
\int \int_{m+n}\hspace{-13pt} \Sigma [|\Psi^{nm}_{S_{+x}}
(x_{1}, k_{\perp 1}, \lambda_1,} \\
&{\ldots}&{, x_{n_{\bar{q}}}, k_{\perp n_{\bar{q}}},
\lambda_{n_{\bar{q}}}, x_{n_{\bar{q}}+1}, k_{\perp n_{\bar{q}} +1},
\lambda_{n_{\bar{q}}+1}, \ldots, x_n, k_{\perp n}, \lambda_{n};\{g\})|^{2}]}
\end{eqnarray*}
$$ \eqno{(4-10b)} $$

$$
\{\sum_{i=n_{\bar{q}} + 1 (1)}^{n(n_{\bar{q}})}
\delta(x-x_i)(\delta_{\lambda_i , \uparrow} - \delta_{\lambda_i, \downarrow})\}
$$
where,
$$
\int_{n+m}\hspace{-12pt} \Sigma  = \sum_{\{ \bar{f}_1, \cdots,
\bar{f}_{n \bar{q}},f_{n_{\bar{q}}+1},
f_{n \bar{q} + 1}\}}
\sum_{\{\lambda_1,\cdots,\lambda_n\}} \sum_{\{g\}_m}
\int[d^3k_{(\rm{f,\bar{f}})} d^3k_g]_{n+m}
%\int[dx]_n[d^2k_\perp]_n
$$
is the sum over all Fock states with $n$-fermions where at least one
quark (antiquark) carries
longitudinal momentum fraction x.  In eqs. (4-9) for n=1,
eqs. (4-10) gives the light-cone
wavefunction representation of the sum rule that was defined
as the tensor charge in \cite{jaffe}$^*$.
%As stated earlier, since the matrix elements
%needed to compute $h_1(x)$ do not mix Fpock space states
%in the wavefunction, it has a interpretation as
%a density in an appropriate basis.
%In particular one

\bigskip
\noindent {\Large\bf $h_2(x)$}
\medskip

We will now extract the coefficients $c_{n,l}$ from the
matrix element (2-9) for $h_2(x)$.  As a suitable
choice for the proton state we take for the momenta
$P=(P^+,M/2P^+,0,0)$ with polarization along the $+z$-direction
so that $S=(P^+,0,0,0)$.  From (2-9) we see
that $c_{n,l}$ can be identified from
$$
c_{nl}={{\langle PS|\theta^{[+]_n}_l|PS\rangle} \over {2M(P^+)^n}}
\eqno(4-11)
$$
In terms of the general form (4-1), the spin
and momentum terms
$O_s (q,q';\{g\}_k,\{g\}_{k'})$,
and $f_p(q,q',\{g\}_k,\{g\}_{k'})$ that are nonvanishing
are given in table 1.  The nonvanishing Fock space components that
contribute are given in table
2-4 and will be further discussed below.  Each entry in these tables
corresponds to the spin and polarization state shown as well as
the one where they are all flipped.  The diagrams in the tables indicate
how the parton distribution operator contracts with the
partons in the wavefunction.  The lines to the left
(right) contract with the bra (ket) states.  Note that
Fock state amplitude coefficients that differ by one gluon also
differ dimensionally by one mass unit.  In the overlap
integrals for matrix elements of $h_2(x)$ this
difference is balanced on the left-hand side of
eqs. (2-9) by the factor M.

All nonvanishing matrix elements connect Fock states which
differ by one gluon.  Furthermore, in tables
2 and 3, the connected Fock states have the same number of $q$ and $\bar q$
with only one fermion in the two respective states differing
in momenta.  In Table 4, the connected Fock states differ by a single
additional $q\bar q$ pair in one state relative
to the other.  Among the eight $q \bar q g$-operatoor combinations
from (4-11) the two missing in tables 2-4 are the case where
in table 4 one swaps places between $\epsilon$ and $\epsilon^*$.
These two cases lead to contractions which were denoted as "disconnected"
in \cite{mankiewicz}.  As argued by these authors, these cases can be
ignored as the states so formed after contraction
decouple.

Tables (2-4) give the full partonic interpretation
of $h_L(x)$ and thus also $h_2(x)$.  To further clarify
the tensor structure we note the identity,
$$
\bar{\psi} \sigma^{j_{+}} i \gamma^{5} \vec{V}_j^{\perp} \psi =
\bar{\psi}  \gamma^{+} [\vec{V}^{\perp} \times \bar{\gamma}^{\perp} +
i\gamma^5 \vec{\gamma}^{\perp} \cdot \vec{V}^{\perp}] \psi
\eqno (4-12)
$$
where $V^\perp$ is an arbitrary vector with nonvanishing components
only in the xy-plane.  We can therefore
express the matrix elements as,
\begin{eqnarray*}
{c_{n,l}} & {=} & {\frac{1}{4M(P^{+})^{n}} <PS| \bar{\psi} \gamma^{+}
(i\partial^{+})^{l-2} \{ \partial^{+} \bar{A}^{\perp} \times
\bar{\gamma}^{\perp} -
i\gamma^{5} \bar{\gamma}^{\perp} \cdot \partial^{+} \bar{A}^{\perp}  \}} \\
& & {(i\partial^{+})^{n-l} \psi |PS>}
\end{eqnarray*}
$$
\eqno (4-13)
$$
\bigskip
\noindent {\Large\bf Conclusion}
\medskip

Our analysis has given the Fock space light-cone wavefunction
decomposition of $h_1(x)$ and $h_2(x)$.  This unifies the interpretation
of these quantities with the more common structure functions
$f_1(x)$ and $g_1(x)$ as well as $g_2(x)$ from the work in \cite{mankiewicz}.
It is more difficult to describe $h_2(x)$ compared to
$f_1(x), g_1(x)$ or $h_1(x)$, as it is not readily interpreted
as a density.  However, we have see that its interpretation
has a natural explanation within the light-cone wavefunction
formalism.

Our analysis shows that like $g_2(x)$,
$h_2(x)$ contains information about the correlation between
Fock space components differing by one gluon and
one quark-antiquark pair.  Furthermore, it has a nonperturbative dependence
on the QCD coupling constant.  By this we mean, the
coupling constant that enters in the matrix elements
tabulated in tables 2-4 is evaluated at its low energy scale.
Experimental information combined with perturbative QCD scaling
formulas, would therefore allow one to deduce nonperturbative
information about low energy quark-gluon correlations.
In the past, low energy constituent quark model descriptions of the hadron
wavefunction have gained considerable insight from experimental
information on $f_1(x)$.  Analogously to gain empirical insight about
gluon correlations which in turn could be used for guidance in forming
low energy models,
experimental data on $g_2(x)$ and $h_2(x)$ would provide
helpful information.

In this paper we have presented the formal building blocks
for the analysis of $h_1(x)$ and $h_2(x)$ in the light-cone
wavefunction formalism.  These results have been of a technical
nature, but are a necessary first step in understanding
the partonic content of these distributions.  We have not
addressed the more pragmatic question of how
to compute them from models.  Let us therefore briefly turn to
this issue.  For $h_1(x)$ it would be interesting to test SU(6)-breaking
quark model wavefunctions such as those in \cite{yau}
and their extensions along the lines of \cite{dzie}.
Turning to $h_2(x)$, one possibility in a low-energy model would
be to expand the constituent quark model hilbert space by the inclusion
of just one additional gluon component.  Such an approach
has already be tried by Lipkin \cite{lipkin} with some
success in explaining the nucleon longitudinal
spin content.  Further insight on modeling "valence gluons"
could be gained from the flux-tube model
of Isgur and Paton \cite{isgur}.
%-------------------------------------

\bigskip
\noindent{\Large\bf Acknowledgements}
\medskip

Financial support was provided by
the Director, Office of Energy Research, Office of High Energy
and Nuclear Physics,
Division of High Energy Physics of the US Dept. of Energy.

\medskip
**  Present address: Department of Physics, the Pennslyvania State University,
104 Davey Laboratory, University Park, PA 16802-6300

$^*$ The expressions for the quark densities in section
3 of \cite{jaffe} are dimensionally inconsistent but can be
easily fixed by insertion of $\delta$-function factors similar
to what is done in eq. (4-10a).

\bigskip
\noindent {\Large\bf Appendix}
\medskip

In all our work, we used the Dirac representation of the
$\gamma$-matrices.
It is often useful in calculations of higher twist distributions
to have readily available, explicit expressions
for the spinors.  For convenience, we have reproduced below
the spinors relevant to our work.
The light-cone Dirac spinors are,

$$ u_{\uparrow}(k) = \frac{1}{\sqrt{2k^+}}\left(
\begin{array}{cc}
k^+ + m \\ k_x + ik_y \\ k^+ - m \\ k_x + ik_y
\end{array}
\right)
\eqno (A-1a)
$$ $$ u_{\downarrow}(k) = \frac{1}{\sqrt{2k^+}}\left(
\begin{array}{cc}
-k_x + ik_y\\ k^+ + m \\ k_x - ik_y \\ -k^+ + m
\end{array}
\right)
\eqno (A-1b)
$$ $$ u_{\uparrow}(k) = \frac{1}{\sqrt{2k^+}}\left(
\begin{array}{cc}
-k_x + ik_y\\ k^+ - m \\ k_x - ik_y \\ -k^+ - m
\end{array}
\right)
\eqno (A-1c)
$$ $$ u_{\downarrow}(k) = \frac{1}{\sqrt{2k^+}}\left(
\begin{array}{cc}
k^+ - m \\ k_x + ik_y \\ k^+ + m \\ k_x + ik_y
\end{array}
\right)
\eqno (A-1d)
$$
For the light-cone spinors in the eigenstate basis
of $Q^i_{+-}$ we have,
The transverse light cone spinors, $u^{i}_{+}(k, \lambda),
v^{i}_{+}(k, \lambda)$ which satisfy (4-4) are,
where below $u_+(k, \lambda), v_+(k, \lambda)$
are the good components of the light-cone spinors.,
\begin{eqnarray*}
{u^{x}_{+}(k, \lambda = +)} & { = } & {u_{+}(k,\lambda= \uparrow) - u_{+}(k,
\lambda=\downarrow)} \\
                            & { = } & {\sqrt{\frac{k^+}{2}}} \left(
\begin{array}{cc}
{ 1} \\ {-1} \\ { 1} \\ { 1}
\end{array}
\right)
\end{eqnarray*}
$$
\eqno{(A-2a)}
$$
\begin{eqnarray*}
{v^{x}_{+}(k, \lambda = +)} & { = } & {v_{+}(k, \lambda = \downarrow) -
v_{+}(k, \lambda = \uparrow)} \\
                            & { = } & {\sqrt{\frac{k^+}{2}}} \left(
\begin{array}{cc}
{1} \\ {-1} \\ {1} \\ {1}
\end{array}
\right)
\end{eqnarray*}
$$
\eqno{(A-2b)}
$$
\begin{eqnarray*}
{u^{x}_{+}(k, \lambda = -)} & { = } & {u_{+}(k,\lambda= \uparrow) - u_{+}(k,
\lambda=\downarrow)} \\
                            & { = } & {\sqrt{\frac{k^+}{2}}} \left(
\begin{array}{cc}
{1} \\ {1} \\ {1} \\ {-1}
\end{array}
\right)
\end{eqnarray*}
$$
\eqno{(A-2c)}
$$
\begin{eqnarray*}
{v^{x}_{-}(k, \lambda = +)} & { = } & {v_{+}(k, \lambda = \downarrow) -
v_{+}(k, \lambda = \uparrow)} \\
                            & { = } & {\sqrt{\frac{k^+}{2}}} \left(
\begin{array}{cc}
{1} \\ {1} \\ {1} \\ {-1}
\end{array}
\right)
\end{eqnarray*}
$$
\eqno{(A-2d)}
$$
\begin{eqnarray*}
{u^{y}_{+}(k, \lambda = +)} & { = } & {u_{+}(k,\lambda = +) - iu_{+}(k,
\lambda=\downarrow)} \\
                            & { = } & {\sqrt{\frac{k^+}{2}}} \left(
\begin{array}{cc}
{ 1} \\ {i} \\ { 1} \\ {-i}
\end{array}
\right)
\end{eqnarray*}
$$
\eqno{(A-2e)}
$$
\begin{eqnarray*}
{v^{y}_{+}(k, \lambda = +)} & { = } & {v_{+}(k, \lambda = \downarrow) -
iv_{+}(k, \lambda = \uparrow)} \\
                            & { = } & {\sqrt{\frac{k^+}{2}}} \left(
\begin{array}{cc}
{1} \\ {i} \\ {1} \\ {-i}
\end{array}
\right)
\end{eqnarray*}
$$
\eqno{(A-2f)}
$$
\begin{eqnarray*}
{u^{y}_{+}(k, \lambda = -)} & { = } & {u_{+}(k,\lambda = +) - iu_{+}(k,
\lambda=\downarrow)} \\
                            & { = } & {\sqrt{\frac{k^+}{2}}} \left(
\begin{array}{cc}
{ 1} \\ {-i} \\ { 1} \\ { i}
\end{array}
\right)
\end{eqnarray*}
$$
\eqno{(A-2g)}
$$
\begin{eqnarray*}
{v^{y}_{+}(k, \lambda = -)} & { = } & {v_{+}(k, \lambda = \downarrow) -
iv_{+}(k, \lambda = \uparrow)} \\
                            & { = } & {\sqrt{\frac{k^+}{2}}} \left(
\begin{array}{cc}
{1} \\ {-i} \\ {1} \\ {i}
\end{array}
\right)
\end{eqnarray*}
$$
\eqno{(A-2h)}
$$

\eject
\centerline{\u{Table Captions}}
Table 1: Momentum and spin terms arising in the Fock space matrix elements
of $h_2(x)$.

Table 2: Evaluation of spin factor for the gluon annihilation
matrix elements to $h_2(x)$.
Table gives expressions for the spin states shown and the case when
all spins are flipped.

Table 3: Evaluation of spin factor for the gluon creation
matrix elements to $h_2(x)$.
Table gives expressions for the spin states shown and the case when
all spins are flipped.

Table 4: Evaluation of spin factor for the quark pair creation and
annihilation
matrix elements to $h_2(x)$.
Table gives expressions for the spin states shown and the case when
all spins are flipped.

\eject

\begin{tabular}{|l|l|l|} \hline
\multicolumn{3}{|c|}{Table 1} \\ \hline
{$O^{n,l}_{s}(q, q', s, s', \{g\})$} & {$f^{n,l}_{p}
(q, q', \{g\})/M(P^{+})^{n-1}$}
& {Process} \\ \hline
{${\bar{u}(k_q)\over\sqrt{x_q}}\gamma^+ \gamma^5 \epsilon
\hspace{-0.1in}/(k_g){u(k_{q'}
= k_q - k_g)\over\sqrt{x_{q'}}}$}
& {$(x_q)^{l-2}x_g(x_q-x_g)^{n-l}$}
& {\begin{picture}(80,80)(0,0)
\put(0,50){\line(1,0){35}}
\put(35,50){\line(3,-2){30}}
\thicklines
\put(35,50){\line(3,2){30}}
\thinlines
\put(15,40){$q$}
\put(45,65){$g$}
\put(45,25){$q'$}
\end{picture}} \\ \hline
{${\bar{v}(k_q)\over\sqrt{x_q}}\gamma^+ \gamma^5 \epsilon
\hspace{-0.1in}/(k_g){v(k_{q'}
= k_q + k_g)\over\sqrt{x_{q'}}}$}
& {$(x_q)^{l-2}x_g(x_q+x_g)^{n-l}$}
& {\begin{picture}(80,80)(0,0)
\put(0,50){\line(1,0){35}}
\put(35,50){\line(3,-2){30}}
\thicklines
\put(35,50){\line(3,2){30}}
\thinlines
\put(15,40){$q'$}
\put(45,65){$g$}
\put(45,25){$q$}
\end{picture}} \\ \hline
{${\bar{u}(k_q)\over\sqrt{x_q}}\gamma^+ \gamma^5 \epsilon \hspace{-0.1in}/^{*}
(k_g){u(k_{q'}
= k_q + k_g)\over\sqrt{x_{q'}}}$}
& {$(x_q)^{l-2}x_g(x_q+x_g)^{n-l}$}
& {\begin{picture}(80,80)(0,0)
\put(30,50){\line(1,0){35}}
\thicklines
\put(0,70){\line(3,-2){30}}
\thinlines
\put(0,30){\line(3,2){30}}
\put(45,40){$q'$}
\put(15,65){$g$}
\put(15,25){$q$}
\end{picture}} \\ \hline
{${\bar{v}(k_q)\over\sqrt{x_q}}\gamma^+ \gamma^5 \epsilon
\hspace{-0.1in}/^{*}(k_g){v(k_{q'}
= k_g - k_q)\over\sqrt{x_{q'}}}$}
& {$(x_q)^{l-2}x_g(x_g-x_q)^{n-l}$}
& {\begin{picture}(80,80)(0,0)
\put(30,50){\line(1,0){35}}
\thicklines
\put(0,70){\line(3,-2){30}}
\thinlines
\put(0,30){\line(3,2){30}}
\put(45,40){$q$}
\put(15,65){$g$}
\put(15,25){$q'$}
\end{picture}} \\ \hline
{${\bar{u}(k_q)\over\sqrt{x_q}}\gamma^+ \gamma^5 \epsilon
\hspace{-0.1in}/(k_g){v(k_{q'}
= k_q - k_g)\over\sqrt{x_{q'}}}$}
& {$(x_q)^{l-2}x_g(x_g-x_q)^{n-l}$}
& {\begin{picture}(80,80)(0,0)
\thicklines
\put(30,50){\line(1,0){35}}
\thinlines
\put(0,70){\line(3,-2){30}}
\put(0,30){\line(3,2){30}}
\put(45,40){$g$}
\put(15,65){$q$}
\put(15,25){$q'$}
\end{picture}} \\ \hline
{${\bar{v}(k_q)\over\sqrt{x_q}}\gamma^+ \gamma^5 \epsilon
\hspace{-0.1in}/(k_g){u(k_{q'}
= k_g - k_q)\over\sqrt{x_{q'}}}$}
& {$(x_q)^{l-2}x_g(x_g-x_q)^{n-l}$}
& {\begin{picture}(80,80)(0,0)
\thicklines
\put(0,50){\line(1,0){35}}
\thinlines
\put(35,50){\line(3,-2){30}}
\put(35,50){\line(3,2){30}}
\put(15,40){$g$}
\put(45,65){$q$}
\put(45,25){$q'$}
\end{picture}} \\ \hline
\end{tabular}
\bigskip

%\noindent Table 2.
%\medskip
\begin{center}
\begin{tabular}{|c|c|c|c|c|}
\hline
\multicolumn{5}{|c|}{Table 2} \\ \hline
\multicolumn{3}{|c|}{Process}
& {\begin{picture}(80,80)(0,0)
\put(0,50){\line(1,0){35}}
\put(35,50){\line(3,-2){30}}
\thicklines
\put(35,50){\line(3,2){30}}
\thinlines
\put(15,40){$k_{q'}$}
\put(45,65){$k_{g}$}
\put(45,25){$k_{q}$}
\end{picture}}
& {\begin{picture}(80,80)(0,0)
\put(0,50){\line(1,0){35}}
\put(35,50){\line(3,-2){30}}
\thicklines
\put(35,50){\line(3,2){30}}
\thinlines
\put(15,40){$k_{q}$}
\put(45,65){$k_{g}$}
\put(45,25){$k_{q'}$}
\end{picture}} \\ \hline

$\lambda_{q^\prime(\bar q^\prime)}$&$\lambda_{q(\bar q)}$&$\lambda_g$&
${1\over p^+}{\bar u_{\lambda_{q^\prime}}(k_{q^\prime})
\over\sqrt{x_{q^\prime}}}\gamma^+\gamma^5{\not{\cal E}_{\lambda_g}(k_g)\over
\sqrt{x_g}}{u_{\lambda_q}(k_q)\over\sqrt{x_q}}$ &
${1\over p^+}{\bar v_{\lambda_{q^\prime}}(k_{q^\prime})
\over\sqrt{x_{q^\prime}}}\gamma^+\gamma^5{\not{\cal E}_{\lambda_g}(k_g)\over
\sqrt{x_g}}{v_{\lambda_q}(k_q)\over\sqrt{x_q}}$  \\
\hline \hline
$\uparrow$& $\downarrow$ & $+1$ & $-2\sqrt{2\over x_{q^\prime}-x_q}$ & $0$ \\
\hline
$\uparrow$ & $\downarrow$ & $-1$ & $0$ & $-2\sqrt{2\over x_{q^\prime}-x_q}$ \\
\hline
\end{tabular}
\end{center}

%\noindent Table 3.
%\medskip
\begin{center}
\begin{tabular}{|c|c|c|c|c|}
\hline
\multicolumn{5}{|c|}{Table 3} \\ \hline
\multicolumn{3}{|c|}{Process}
&{\begin{picture}(80,80)(0,0)
\put(30,50){\line(1,0){35}}
\thicklines
\put(0,70){\line(3,-2){30}}
\thinlines
\put(0,30){\line(3,2){30}}
\put(45,40){$k_{q}$}
\put(15,65){$k_{g}$}
\put(15,25){$k_{q'}$}
\end{picture}}
&{\begin{picture}(80,80)(0,0)
\put(30,50){\line(1,0){35}}
\thicklines
\put(0,70){\line(3,-2){30}}
\thinlines
\put(0,30){\line(3,2){30}}
\put(45,40){$k_{q'}$}
\put(15,65){$k_{g}$}
\put(15,25){$k_{q}$}
\end{picture}} \\ \hline

$\lambda_{q^\prime(\bar q^\prime)}$&$\lambda_{q(\bar q)}$&$\lambda_g$&
${1\over p^+}{\bar u_{\lambda_{q^\prime}}(k_{q^\prime})
\over\sqrt{x_{q^\prime}}}\gamma^+\gamma^5{\not{\cal E}_{\lambda_g}^*(k_g)\over
\sqrt{x_g}}{u_{\lambda_q}(k_q)\over\sqrt{x_q}}$ &
${1\over p^+}{\bar v_{\lambda_{q^\prime}}(k_{q^\prime})
\over\sqrt{x_{q^\prime}}}\gamma^+\gamma^5{\not{\cal E}_{\lambda_g}^*(k_g)\over
\sqrt{x_g}}{v_{\lambda_q}(k_q)\over\sqrt{x_q}}$  \\
\hline \hline
$\uparrow$& $\downarrow$ & $-1$ & $-2\sqrt{2\over x_{q^\prime}-x_q}$ & $0$ \\
\hline
$\uparrow$ & $\downarrow$ & $+1$ & $0$ & $-2\sqrt{2\over x_{q^\prime}-x_q}$ \\
\hline
\end{tabular}
\end{center}

%\noindent Table 4.
%\medskip
\begin{center}
\begin{tabular}{|c|c|c|c|c|}
\hline
\multicolumn{5}{|c|}{Table 4} \\ \hline
\multicolumn{3}{|c|}{Process}
& {\begin{picture}(80,80)(0,0)
\thicklines
\put(30,50){\line(1,0){35}}
\thinlines
\put(0,70){\line(3,-2){30}}
\put(0,30){\line(3,2){30}}
\put(45,40){$k_{g}$}
\put(15,65){$k_{q}$}
\put(15,25){$k_{q'}$}
\end{picture}}
& {\begin{picture}(80,80)(0,0)
\thicklines
\put(0,50){\line(1,0){35}}
\thinlines
\put(35,50){\line(3,-2){30}}
\put(35,50){\line(3,2){30}}
\put(15,40){$k_{g}$}
\put(45,65){$k_{q}$}
\put(45,25){$k_{q'}$}
\end{picture}} \\ \hline

$\lambda_{q}$&$\lambda_{\bar q}$&$\lambda_g$&
${1\over p^+}{\bar u_{\lambda_{q^\prime}}(k_{q^\prime})
\over\sqrt{x_{q^\prime}}}\gamma^+\gamma^5{\not{\cal E}_{\lambda_g}(k_g)\over
\sqrt{x_g}}{v_{\lambda_q}(k_q)\over\sqrt{x_q}}$ &
${1\over p^+}{\bar v_{\lambda_{q^\prime}}(k_{q^\prime})
\over\sqrt{x_{q^\prime}}}\gamma^+\gamma^5{\not{\cal E}_{\lambda_g}^*(k_g)\over
\sqrt{x_g}}{u_{\lambda_q}(k_q)\over\sqrt{x_q}}$  \\
\hline \hline
$\uparrow$& $\downarrow$ & $-1$ & $-2\sqrt{2\over x_{q^\prime}-x_q}$ &
$-2\sqrt{2\over x_{q^\prime}-x_q}$ \\
\hline
\end{tabular}
\end{center}

\end{document}